\documentclass{pasj00}
\begin{document}
\SetRunningHead{J.\ Fukue}
{Relativistic Variable Eddington Factor}
%in Relativistic Radiative Flow}
\Received{yyyy/mm/dd}
\Accepted{yyyy/mm/dd}
%tex          2007 1209
%referee      2008 0321
%editing      

\title{Variable Eddington Factor in a Relativistic Plane-Parallel Flow}
%in Relativistic Radiative Flow}

%%% begin:list of authors
\author{Jun \textsc{Fukue}} %   \thanks{}}
\affil{Astronomical Institute, Osaka Kyoiku University, 
Asahigaoka, Kashiwara, Osaka 582-8582}
\email{fukue@cc.osaka-kyoiku.ac.jp}

%\author{B-Firstname \textsc{B-Familyname}}
%\affil{B-Address of Institute}\email{bbbbb@xxx.xxx.xx.xx}
%\and
%\author{C-Firstname {\sc C-Familyname}}
%\affil{C-Address of Institute}\email{ccccc@xxx.xxx.xx.xx}
%%% end:list of authors

%% `\KeyWords{}' always has to be placed before `\maketitle'.
 %Do NOT move this preamble from here!

\maketitle

%\newpage

\begin{abstract}
We examine the Eddington factor in an optically thick,
relativistic flow accelerating in the vertical direction.
When the gaseous flow is radiatively accelerated 
and there is a velocity gradient,
there also exists a density gradient.
The comoving observer sees radiation coming from
a closed surface where the optical depth measured from the observer is unity.
Such a surface, called a {\it one-tau photo-oval},
 is elongated in the flow direction.
In general, the radiation intensity emitted by the
photo-oval is non-uniform, and the photo-oval surface has a
relative velocity with respect to the position of the comoving observer.
Both effects introduce some degree of anisotropy in the radiation field
observed in the comoving frame.
As a result, the radiation field observed by the comoving observer
becomes {\it anisotropic}, 
and the Eddington factor must deviate from the usual value of 1/3.
Thus,
the relativistic Eddington factor generally depends
on the optical depth $\tau$
and the velocity gradient $du/d\tau$,
$u$ being the four velocity.
In the case of a plane-parallel vertical flow,
we obtain the shape of the photo-oval and 
calculate the Eddington factor
in the optically thick regime.
We found that the Eddington factor $f$
is well approximated by
$f(\tau, \frac{du}{d\tau}) = 
\frac{1}{3} 
\exp ( \frac{1}{u}
\frac{du}{d\tau} ) $.
%Using the analytical result of the subrelativistic regime,
%the suitable form in all relevant range would be
%$f = \frac{1}{3}
%\exp ( \frac{16}{15} 
%\frac{\beta}{u} 
%\frac{du}{d\tau} ) $.
%
%$f(\tau, u, \frac{du}{d\tau}) = 
%\frac{\displaystyle 1}{\displaystyle 3} 
%\exp \left( \frac{\displaystyle 1}{\displaystyle u}
%\frac{\displaystyle du}{\displaystyle d\tau} \right) $.
%Using the analytical result of the subrelativistic regime,
%the suitable form in all relevant range would be
%$f = \frac{\displaystyle 1}{\displaystyle 3}
%\exp \left( \frac{\displaystyle 16}{\displaystyle 15} 
%\frac{\displaystyle \beta}{\displaystyle u} 
%\frac{\displaystyle du}{\displaystyle d\tau} \right) $.
%
This relativistic variable Eddington factor can be used
in various relativistic radiatively-driven flows.
%such as black-hole accretion flows,
%relativistic astrophysical jets and outflows, and
%relativistic explosions like gamma-ray bursts.
\end{abstract}

\section{Introduction}

In the moment formalism of radiation hydrodynamics,
in order to close the moment equations truncated at a finite order,
{\it closure relations},
such as the Eddington approximation or more complex expressions,
are usually adopted.
In a non-relativistic static atmosphere
(Chandrasekhar 1960; Mihalas 1970; Rybicki \& Lightman 1979;
Mihalas \& Mihalas 1984; Shu 1991; Peraiah 2002; Castor 2004),
for example,
many researchers have used the Eddington approximation
and variable Eddington factors as closure relations
(Milne 1921; Eddington 1926; Kosirev 1934; Chandrasekhar 1934;
Hummer \& Rybicki 1971; Wilson et al. 1972; Tamazawa et al. 1975;
Unno \& Kondo 1976; Masaki \& Unno 1978; ).
In relativistically moving flows
(Mihalas et al. 1975; Mihalas et al. 1976a, b;
Thorne 1981; Thorne et al. 1981; Flammang 1982, 1984;
Mihalas \& Mihalas 1984; Nobili et al. 1991, 1993; 
Kato et al. 1998, 2008; Castor 2004; Mihalas \& Auer 2001;
Park 2001, 2006; Takahashi 2007),
the Eddington approximation is written in the comoving frame,
and then transformed to the inertial frame, if necessary
(Castor 1972; Hsieh \& Spiegel 1976; Fukue et al. 1985;
Sen \& Wilson 1993; Baschek et al. 1995, 1997; Kato et al. 1998, 2008).

As is well known, the standard Eddington approximation is valid
when the radiation field is almost {\it isotropic}.
This conditions is violated in the optically thin regime
or the relativistic regime with strong velocity gradients.
%Actually, the pathological behavior
%in relativistic radiation moment equations
%has been pointed out and examined
%(Turolla \& Nobili 1988; Nobili et al. 1991; 
%Turolla et al. 1995; Dullemond 1999).
%Namely, the moment equations for radiation transfer
%in relativistically moving media
%can generally have singular (critical) points.
%As a result, solutions behave pathologically
%in a relativistic regime.
%The appearance of singularities is supposed to be
%related to the approximation of the full transfer equations
%with a finite number of moments (Dullemond 1999).
Furthermore, in a relativistically accelerating flow, 
even for a comoving observer,
the comoving radiation field may become {\it anisotropic}.
In order to improve the situation we are confronted with,
instead of the usual Eddington approximation,
one should adopt a {\it variable Eddington factor},
which depends on the velocity gradient $d\beta/d\tau$
as well as the optical depth $\tau$
(Fukue 2006; Fukue \& Akizuki 2006, 2007; Akizuki \& Fukue 2008;
Fukue 2008; Koizumi \& Umemura 2008).

In this paper
we semi-analytically consider and derive a suitable form for
the variable Eddington factor for a relativistic, plane-parallel,
scattering medium,
using a simple but physically consistent treatment
of the radiation field in the optically thick regime.

In the next section
we describe the shape of the surface
where the optical depth measured by the comoving observer is unity,
and in section 3 we derive its expression 
when the four-velocity gradient is large.
In section 4
we numerically calculate
the comoving radiation field within the photo-oval,
and the relativistic variable Eddington factor.
In section 5,
we briefly discuss the present result and compare with the results
of previous work.
The final section is devoted to concluding remarks.

\section{One-Tau Photo-Oval in the Comoving Frame}

Let us suppose an optically thick, relativistic radiative flow,
which is accelerated in the vertical ($z$) direction,
and a comoving observer,
who moves upward with the flow,
at an optical depth $\tau$ (figure 1).

\begin{figure}
  \begin{center}
  \FigureFile(70mm,70mm){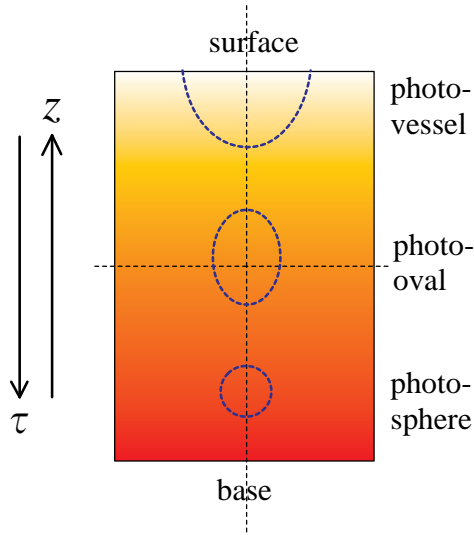}
  \end{center}
\caption{
Schematic picture of a relativistic radiative flow
in the vertical direction.
The flow is accelerated in the vertical ($z$) direction,
and has a velocity gradient.
The dashed curves are
one-tau photo-ovals observed by a comoving observer.
}
\end{figure}

Sufficiently deep inside the flow,
where the optical depth is very large,
the mean free path of photons is very short.
In such a case,
within the range of the mean free path,
for the comoving observer
the flow is seen to be almost uniform;
the velocity gradient and resultant density gradient are negligible.
Hence,
the mean free path is the same in all directions,
and the shape of the surface
where the optical depth measured from the comoving observer is unity,
is almost a sphere;
we call it a {\it one-tau photo-sphere} (a dashed circle in figure 1).
As a result,
the comoving radiation field is nearly isotropic,
and a usual Eddington approximation in the comoving frame is valid.

On the other hand,
if the velocity gradient becomes very large,
the behavior of the radiation field may be affected
by changes in the hydrodynamics of the flow
even on a lengthscale comparable to that 
of the photon mean free path.
%In the region
%where the velocity is large and the density is low,
%or
%where the velocity gradient is large
%in spite of the high density,
%within the range of the mean free path,
%the velocity gradient cannot be neglected,
%and the density is no longer uniform
%even in the comoving frame of the gas.
Hence,
the mean free path becomes longer in the downstream direction
than in the upstream and other directions, and
the shape of the one-tau range elongates in the downstream direction;
we call it a {\it one-tau photo-oval} (a dashed oval in figure 1).
As a result,
the comoving radiation field becomes {\it anisotropic},
and we should modify the usual Eddington approximation;
in these conditions the Eddington factor may depend on 
the optical depth and the velocity gradient.

In addition,
when the optical depth is sufficiently small
and/or the velocity gradient is suffiently large,
the mean free path of photons in the downstream direction
become less than unity.
Hence,
the shape of the one-tau range is open in the downstream direction;
we call it a {\it one-tau photo-vessel} (a dashed hemi-circle in figure 1).

In order to obtain an appropriate form
of the relativistic variable Eddington factor in these regimes,
we thus carefully treat and examine
the radiation field in the comoving frame.

In a previous paper (Fukue 2008, hereafter refered to as paper I), 
we have examined the one-tau photo-oval,
derived the variable Eddington factor
in the non-relativistic regime, and proved that
the Eddington factor decreases with the velocity gradient.
In paper I, however, 
the treatment was restricted to the non-relativistic regime
of $\beta \leq 0.1$,
where $\beta$ ($=v/c$) is the flow velocity $v$
normalized to the speed of light.

In the present study,
we thus examine the behavior of the relativistic variable Eddington factor
in the relativistic regime of $\beta \sim 1$,
under the same approach of paper I.
Instead of $\beta$, however,
we use the four velocity $u$ ($=\gamma \beta$),
where $\gamma$ ($=1/\sqrt{1-\beta^2}$) is the Lorentz factor,
and seek the appropriate form of $f(\tau, du/d\tau)$.

\begin{figure}
  \begin{center}
  \FigureFile(80mm,80mm){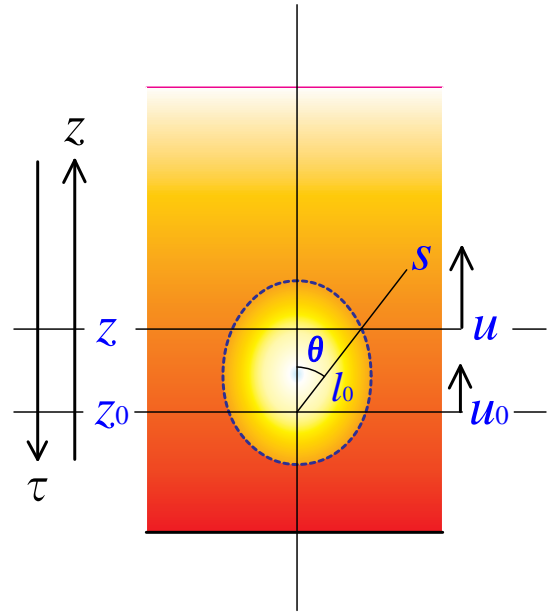}
  \end{center}
\caption{
One-tau photo-oval around a comoving observer
in the vertical ($z$) one-dimensional radiative flow.
The comoving observer is located at $z=z_0$,
where the flow four speed is $u=u_0$ and the optical depth is $\tau=\tau_0$.
In the $s$-direction,
that forms an angle $\theta$ with the downstream direction,
the mean free path is set to be $l_0$.
}
\end{figure}

%%%%%%%%%%%%%%%%%%%%%%%%%%%%%%%%%%%%%%%%%%

\section{Shape of the One-Tau Photo-Oval in the Optically Thick Regime}

In this section
we compute the shape of the one-tau photo-oval
in a vertical, optically thick one-dimensional radiative flow
under the linear approximation.

We define the one-tau photo-oval as the surface
where the optical depth measured by the comoving observer is unity.
The situation is schematically illustrated in figure 2.
We assume that the comoving observer in the vertical flow is located at $z=z_0$,where the flow four speed is $u=u_0$
and the optical depth is $\tau=\tau_0$.
In the $s$-direction,
that forms an angle $\theta$ with the downstream direction,
the mean free path of photons is $l_0$.
The relation among these quantities is
\begin{equation}
   z - z_0 = s \cos\theta.
\label{zz0}
\end{equation}

The continuity equation for the stationary, one-dimensional 
relativistic flow is
\begin{equation}
   \rho cu = J ~(={\rm const.}),
\label{rho}
\end{equation}
where $\rho$ is the proper gas density,
$u$ the four velocity,
and $J$ the mass-flow rate per unit area.
In addition, the optical depth $\tau$ is defined by
\begin{equation}
   d\tau \equiv -\kappa \rho dz,
\label{tau}
\end{equation}
where $\kappa$ is the opacity,
which we assume constant in the present analysis.

In this paper
we use a linear approximation for the flow field;
that is to say, around the position of the comoving observer
the flow four speed is expanded as
\begin{equation}
   u = u_0 + \left.\frac{du}{dz}\right|_0 ~(z-z_0).
\label{linear_u}
\end{equation}

It should be noted that
in paper I
we expanded the flow speed $\beta$,
and derived several results analytically,
restricting the treatment to the non-relativistic regime.
Instead, in the present paper
we expand the four velocity $u$,
and examine the relativistic regime,
although solutions must be calculated numerically.

Under these assumptions,
we consider two different regimes, as outlined below.

\subsection{Linear Regime}

We first consider the linear regime,
where the density is expanded as
\begin{eqnarray}
   \rho &=& \rho_0 + \left.\frac{d\rho}{dz}\right|_0 ~(z-z_0)
\nonumber \\
        &=& \rho_0 + \left.\frac{d\rho}{du}\frac{du}{dz}\right|_0 ~(z-z_0)
\nonumber \\
        &=& \rho_0 - \left.\frac{\rho}{u}\frac{du}{dz}\right|_0 ~(z-z_0),
\label{linear_rho}
\end{eqnarray}
where we made use of the continuity equation (\ref{rho}).

Then the optical depth $\tau_s$ along the $s$-direction
is easily calculated as
\begin{eqnarray}
   \tau_s &=& \int_0^{l_0} \kappa\rho ds
\nonumber \\
%          &=& \int_0^{l_0} \kappa \left[ \rho_0 + \left.\frac{d\rho}{du}\frac{du}{dz}\right|_0 ~(z-z_0) \right] ds
%\nonumber \\
%          &=& \int_0^{l_0} \kappa \left[ \rho_0 + \left.\frac{d\rho}{du}\frac{du}{dz}\right|_0 ~s \cos\theta \right] ds
%\nonumber \\
          &=& \kappa\rho_0 l_0 + \kappa \left.\frac{d\rho}{du}\frac{du}{dz}\right|_0 \frac{\cos\theta}{2} l_0^2.
\label{linear_tau_s}
\end{eqnarray}
Hence, the length $l_0$ at which $\tau_s=1$ is determined by
the quadratic equation,
\begin{equation}
          \kappa \left.\frac{d\rho}{du}\frac{du}{dz}\right|_0 \frac{\cos\theta}{2} l_0^2 + \kappa\rho_0 l_0 -1 = 0.
\label{linear_l0}
\end{equation}

As is easily seen from equation (\ref{linear_l0}),
if there is no velocity gradient,
$\kappa\rho_0 l_0=1$, and
the shape of the one-tau range is a sphere
(one-tau photo-sphere).

The quadratic equation (\ref{linear_l0}) can be easily solved,
giving
\begin{equation}
     \kappa \rho_0 l_0 = \frac{1-\sqrt{1-2a\cos\theta}}{a\cos\theta},
\label{linear_kapparhol}
\end{equation}
where
\begin{eqnarray}
    a &\equiv& - \frac{1}{u} \left.\frac{du}{d\tau}\right|_0
          = - \left.\frac{d}{d\tau} \ln u \right|_0.
\label{linear_a}
\end{eqnarray}

\begin{figure}
  \begin{center}
  \FigureFile(80mm,80mm){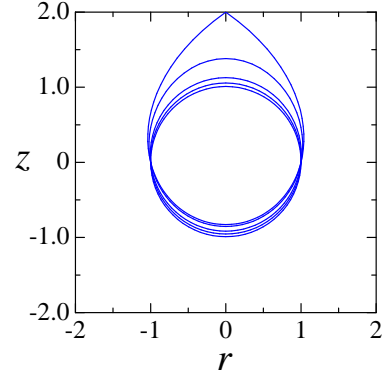}
  \end{center}
\caption{
Shapes of one-tau photo-ovals in the linear regime
for several values of the parameters $a$
(0.02, 0.1, 0.2, 0.4, 0.5), from spherical to elongated.
}
\end{figure}

In figure 3
the solutions (\ref{linear_kapparhol}) 
of the quadratic equation (\ref{linear_l0})
are shown for several values of the parameters $a$
(0.02, 0.1, 0.2, 0.4, 0.5), from spherical to elongated.
For $a>0.5$,
in the downstream direction
there is no solution and the photo-oval
is open, becoming a photo-vessel.
It should be noted that
our figure 3 is equivalent to figure 3 in paper I,
although the definition of the parameter $a$ is different.

 From the solution (\ref{linear_kapparhol}),
the breakup condition, where the photo-oval is open
in the downstream direction, is given by
$1-2a\cos\theta<0$, or
\begin{equation}
      - \left.\frac{1}{u}\frac{du}{d\tau}\right|_0 \cos\theta
      > \frac{1}{2}.
\label{linear_breakup}
\end{equation}
Using the continuity equation (\ref{rho}) and the optical depth (\ref{tau}),
the breakup condition (\ref{linear_breakup}) is also written as
\begin{equation}
      \left.\frac{du}{dz}\right|_0 \cos\theta
      > \frac{\kappa J}{2c}.
\label{linear_breakup2}
\end{equation}
%Hence,
%the condition is somewhat different appearance,
%whether we consider the velocity gradient
%in the linear length $z$ or the optical depth $\tau$.

We have shown that
both the shape of the one-tau photo-oval and its breakup condition
depend on the {\it velocity gradient}.

\subsection{Semi-Linear Regime}

We next consider the semi-linear regime,
where the density is given by continuity equation (\ref{rho}).
When the flow speed is high and the density decreases,
the mean free path of photons becomes large,
and the one-tau range expands.
In such a case,
even if the flow field around the comoving observer
is assumed to be linear in the manner of equation (\ref{linear_u}),
there is no guarantee that, in this regime,
the density distribution can be approximated by equation (5).

Using continuity equation (\ref{rho}),
the optical depth $\tau_s$ along the $s$-direction
is expressed as
\begin{eqnarray}
   \tau_s &=& \int_0^{l_0} \kappa\rho ds
    = \frac{\kappa J}{c} \int_0^{l_0} \frac{1}{u} ds.
\label{semi_tau_s}
\end{eqnarray}
 From equations (\ref{zz0}) and (\ref{linear_u}), we have
\begin{eqnarray}
   u &=& u_0 + \left.\frac{du}{dz}\right|_0 ~s \cos\theta,
\label{semi_u}
\\
   du &=& \left.\frac{du}{dz}\right|_0 ~\cos\theta ~ds,
\end{eqnarray}
and the integral is transformed as
\begin{equation}
   \tau_s = \frac{\kappa J}{c} \frac{1}{\left.\frac{\displaystyle du}{\displaystyle dz}\right|_0 \cos\theta}
              \int_{u_0}^{u} \frac{du}{u},
\label{semi_tau_s2}
\end{equation}
where we use the assumption that the velocity gradient is constant
around $z=z_0$.
This equation (\ref{semi_tau_s2}) is analytically integrated to give
\begin{eqnarray}
   \tau_s &=& \frac{\kappa J}{c} \frac{1}{\left.\frac{\displaystyle du}{\displaystyle dz}\right|_0 \cos\theta}
                  \log \frac{u}{u_0}
          = - \frac{u_0}{\left.\frac{\displaystyle du}{\displaystyle d\tau}\right|_0 \cos\theta}
                  \log \frac{u}{u_0}.
\label{semi_tau_s3}
\end{eqnarray}

Thus, the length $l_0$ at which $\tau_s=1$ 
is finally determined,
with the help of equation (\ref{semi_u})
and the definition of the optical depth (\ref{tau}),
by the following equations:
\begin{eqnarray}
   0 &=&  {\left.\frac{du}{d\tau}\right|_0 \cos\theta}
                + u_0 \log \frac{u}{u_0},
\label{semi_l0}
\\
    u &=& u_0 - {\left.\frac{du}{d\tau}\right|_0 \cos\theta}
                  ~\kappa\rho_0 l_0.
\label{semi_u0}
\end{eqnarray}
We finally obtain
\begin{equation}
\kappa\rho_0 l_0 =
\frac{e^{\left.-\frac{\displaystyle 1}{\displaystyle u}\frac{\displaystyle du}{\displaystyle d\tau}\right|_0 \displaystyle \cos\theta}-1}
     {\left.-\frac{\displaystyle 1}{\displaystyle u}\frac{\displaystyle du}{\displaystyle d\tau}\right|_0 \cos\theta}.
\label{semi_l0final}
\end{equation}

\begin{figure}
  \begin{center}
  \FigureFile(80mm,80mm){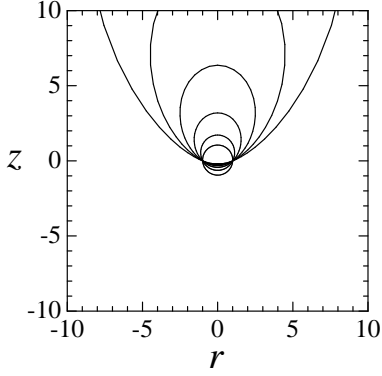}
  \end{center}
\caption{
Shapes of one-tau photo-ovals in the semi-linear regime
for several values of the parameters $a$
(0.1, 1, 2, 3, 4, 5), from spherical to elongated
}
\end{figure}

In figure 4
the one-tau length $\kappa\rho_0 l_0$ (\ref{semi_l0final})
is plotted as a function of $\theta$
for several values of the parameters $a$ ($=-d\ln u/d\tau|_0$)
(0.1, 1, 2, 3, 4, 5), from spherical to elongated.
When the velocity gradient is small,
the photo-oval is spherical.
As the (logarithmic) velocity gradient increases,
it elongates in the downstream direction.
However, in contrast to the linear case
in the previous subsection or in paper I,
for any values of $a$
equation (\ref{semi_l0final}) has always solutions,
and the photo-oval never break up in the downstream direction.

%%%%%%%%%%%%%%%%%%%%%%%%%%%%%%%%%%%%%%%%%%

\section{Comoving Radiation Field and the Relativistic Variable Eddington Factor}

We now derive an expression for the Eddington factor
in the comoving frame assuming that
electron scattering is the dominant radiative process;
the radiation seen by a comoving observer
is that emitted by the one-tau photo-oval around him (figure 5).

In a static and optically thick atmosphere,
the radiation field is isotropic and uniform.
In the present moving atmosphere, on the other hand,
there are two reasons for which the radiation field is not isotropic:
the emitted intensity is not uniform, and
it is redshifted because of velocity gradients.
Neglecting the local emissivity,
the frequency-integrated intensity $I_{\rm co}$ observed in the comoving frame
at the position $z_0$ of the comoving observer is related to 
the frequency-integrated intensity $I$ emitted at the one-tau photo-oval and
to the redshift $z$ by
\begin{equation}
   I_{\rm co} = \frac{I}{(1+z)^4}.
\end{equation}

\begin{figure}
  \begin{center}
  \FigureFile(80mm,80mm){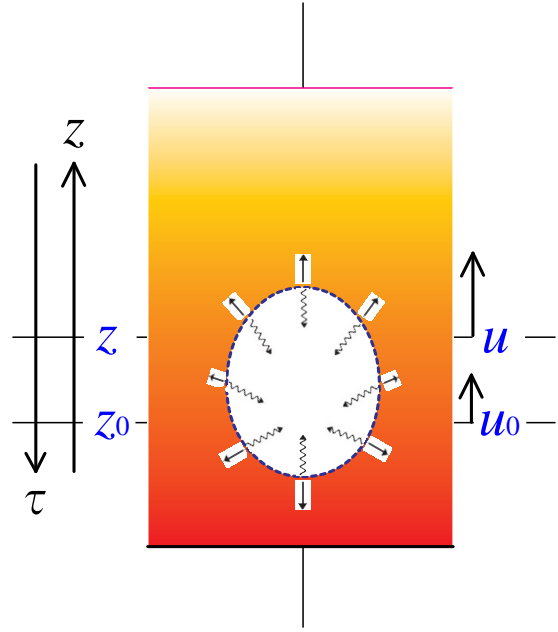}
  \end{center}
\caption{
Radiation field in the one-tau photo-oval around a comoving observer
in the vertical one-dimensional radiative flow.
The radiation field may become anisotropic,
since in general the emitted intensity is not uniform and
it is redshifted due to the velocity difference.
}
\end{figure}

First, the radiative intensity $I$ in the comoving frame
emitted from the one-tau photo-oval walls is not known
without solving the full transfer problem.
Clearly, it is not generally constant,
but is a function of $z$ (or $\tau$) and $\mu$ ($=\cos\theta$).
If the optical depth is sufficiently large,
the $\mu$-dependence may be safely ignored,
and the emitted intensity can be assumed to be isotropic.
In general, however,
the radiative intensity increases with the optical depth,
and therefore,
the intensity from the upstream direction is slightly
larger than that from the downstream direction
in the present one-dimensional vertical flow.
This effect of non-uniformity of intensity
generally acts as a force to accelerate a comoving observer.
As long as the intensity is a slowly increasing function
of the optical depth,
its non-uniformity does not affect significantly
the radiation energy density and the radiation pressure,
and therefore, the Eddington factor (see paper I).
Hence, in order to isolate the effects induced by the relative velocity between
the comoving observer and the one-tau photo-oval,
in this paper we assume that
the intensity of the radiation emitted
 from the one-tau photo-oval surface is constant; $I=I_0$.

Second,
the intensity observed by the comoving observer
is redshifted (Doppler shifted) due to the velocity difference between
the comoving observer and the one-tau photo-oval walls.
In an accelerating flow,
where the flow speed increases toward the downstream direction,
the relative velocity is generally positive (figure 5)
except for some special direction ($\theta=\pi/2$).
Hence, the Doppler shift of intensity also causes anisotropy
of the radiation field at the position of the comoving observer.

The relative speed $\Delta \beta$ between the comoving observer
and the one-tau photo-oval walls is given
by the relativistic summation law as
\begin{equation}
   \Delta \beta = \frac{\beta - \beta_0}{1-\beta\beta_0},
\end{equation}
where
\begin{eqnarray}
   \beta &=& \frac{u}{\sqrt{1+u^2}},
\\
   \beta_0 &=& \frac{u_0}{\sqrt{1+u_0^2}}.
\end{eqnarray}

In addition, using equations (\ref{semi_u0}) and (\ref{semi_l0final}),
the four velocity at $z$ is expressed as
\begin{equation}
   u = u_0 e^{\left.-\frac{\displaystyle 1}{\displaystyle u}\frac{\displaystyle du}{\displaystyle d\tau}\right|_0 \displaystyle \cos\theta}.
\end{equation}
Hence, all the necessary quantities are expressed
as functions of $u_0$, $du/d\tau|_0$, and $\cos\theta$.

Using the relative speed above,
the redshift $z$ is expressed as
\begin{equation}
   1+z = \frac{1+\Delta \beta \cos\theta}{\sqrt{1-(\Delta\beta)^2}},
\end{equation}
and the observed intensity $I_{\rm co}$ is given by equation (20).
%\begin{eqnarray}
%   I_{\rm co} &=& \frac{I_0}{(1+z)^4}.
%\label{RF_Ico}
%\end{eqnarray}
Furthermore, the radiation energy density $E_{\rm co}$,
the radiative flux $F_{\rm co}$, and
the radiation pressure $P_{\rm co}$
measured by the comoving observer are calculated respectively as
\begin{eqnarray}
   cE_{\rm co} &\equiv& \int I_{\rm co} d\Omega_{\rm co},
\label{RF_Eco_linear}
\\
   F_{\rm co} &\equiv& \int I_{\rm co} \cos\theta d\Omega_{\rm co},
\label{RF_Fco_linear}
\\
   cP_{\rm co} &\equiv& \int I_{\rm co} \cos^2\theta d\Omega_{\rm co}.
\label{RF_Pco_linear}
\end{eqnarray}
Finally, the Eddinton factor in the comoving frame is given by
\begin{equation}
   f \equiv \frac{P_{\rm co}}{E_{\rm co}}.
\label{f_linear}
\end{equation}

\begin{figure}
  \begin{center}
  \FigureFile(80mm,80mm){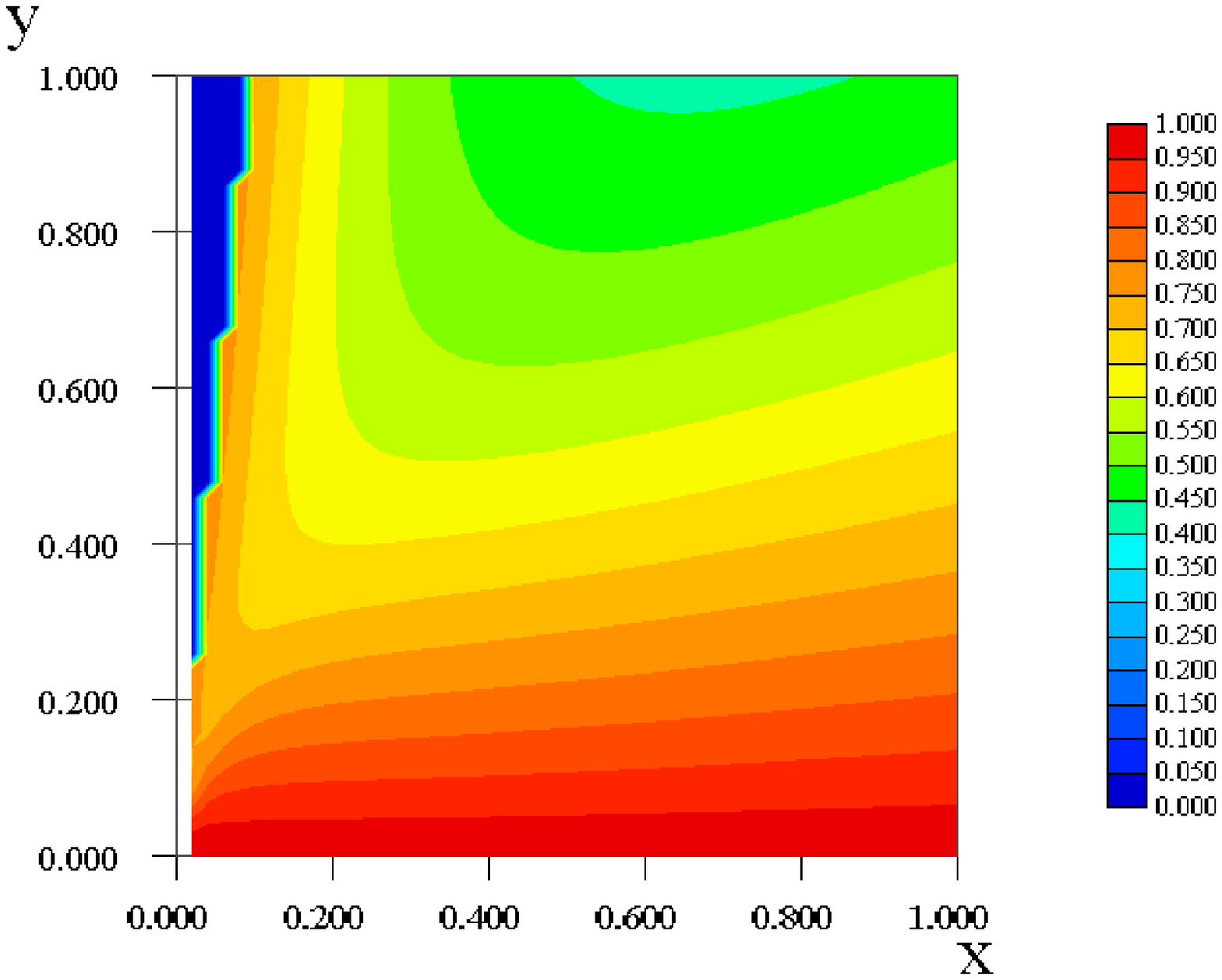}
  \FigureFile(80mm,80mm){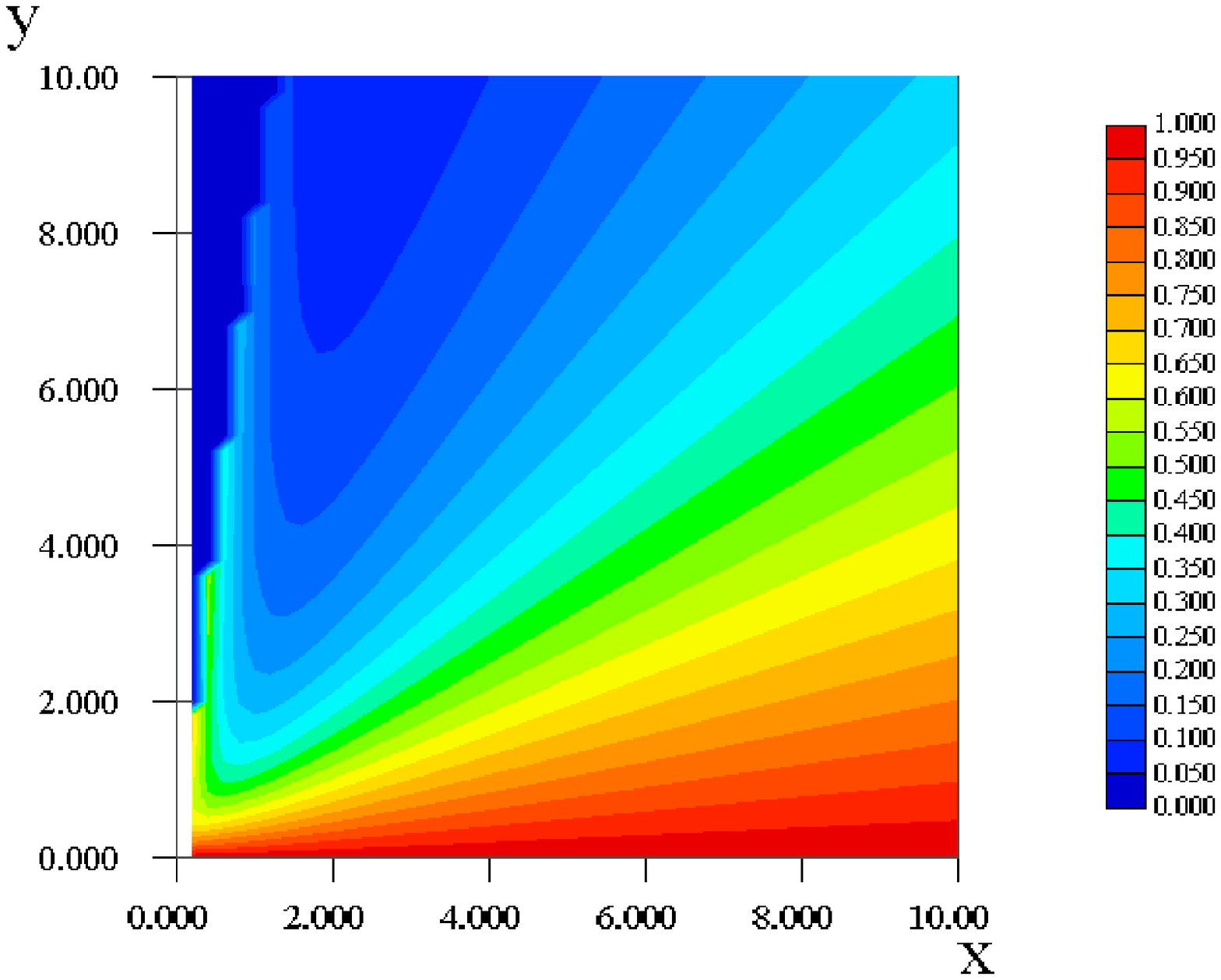}
  \FigureFile(80mm,80mm){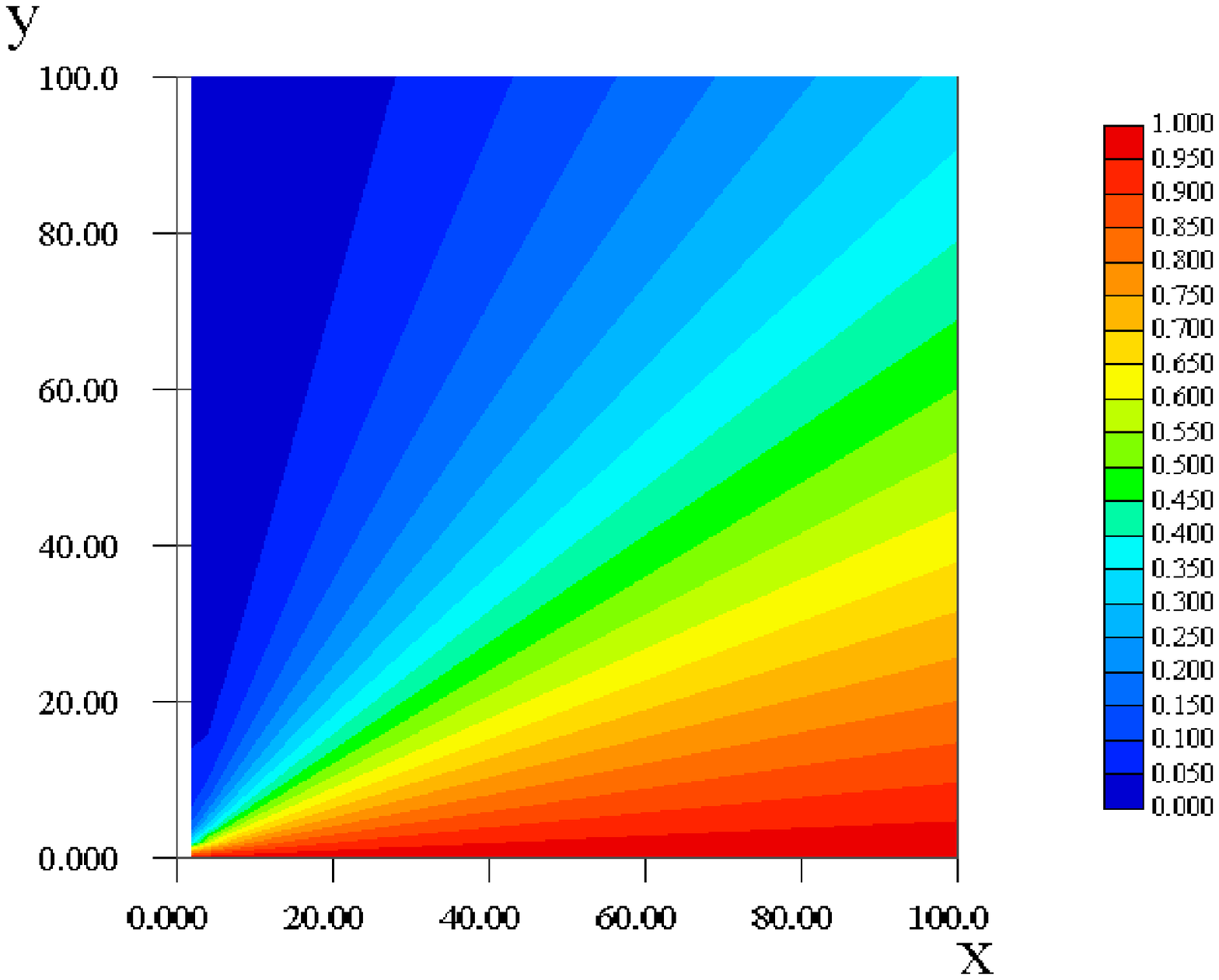}
  \end{center}
\caption{
Contour maps of the relativistic variable Eddington factor $f(u, -du/d\tau)$
multiplied by 3, plotted as a function of the four velocity $u_0$ ($x$ axis)
and the four-velocity gradient $-du/d\tau|_0$ ($y$ axis).
The three panels refer to different intervals of values 
of $u_0$ and $-du/d\tau|_0$.
}
\end{figure}

The numerical results of this calculation are shown in figure 6.
In the non-relativistic regime ($u < 1$) and for small velocity gradients,
the Eddington factor is roughly constant, as expected.
That is, when the velocity gradient is relatively small,
the Eddington factor is almost 1/3.
As the velocity gradient becomes large, however,
the Eddington factor decreases,
as found in paper I.

On the other hand,
the behavior in the highly relativistic regime
is somewhat different.
As can be seen in figure 6,
in the relativistic regime,
the Eddington factor assumes constant values
on the line $-du/d\tau \propto u$.
Considering this behavior, in figure 7
we then plot $3f$
as functions of the four velocity $u_0$ (abscissa)
and the logarithmic velocity gradient $a$ ($=-d\ln u/d\tau|_0$) (ordinate).
As is clearly seen in figure 7,
except for the non-relativistic regime ($u \leq 1$),
the Eddington factor does depend only on
the logarithmic velocity gradient $a$.
We found that
this dependence of the Eddington factor on $a$
is well approximated by
\begin{equation}
   f = \frac{1}{3} e^{-a} = \frac{1}{3} 
   \exp \left( \frac{1}{u}\frac{du}{d\tau} \right).
\label{f}
\end{equation}
In summary,
the relativistic Eddington factor in the vertical flow
varies exponentially
with the logarithmic velocity gradient
in the relativistic regime.

In the present analysis (and paper I)
we have assumed that the comoving intensity is uniform.
Hence, the above result is only due to the redshift effect.
%Indeed, the comoving radiation field
%also decreases as the logarithmic velocity gradient increases
%due to the redshift between the comoving observer
%and the one-tau photo-oval walls.

\begin{figure}
  \begin{center}
  \FigureFile(80mm,80mm){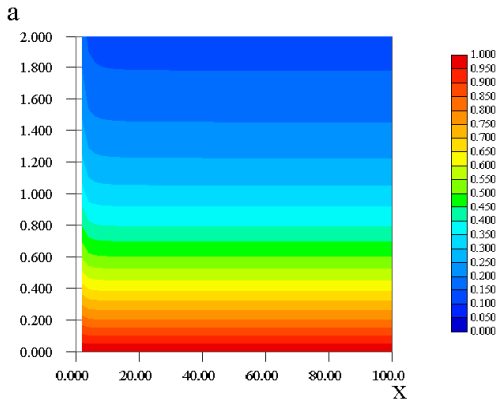}
  \end{center}
\caption{
Contour maps of the relativistic variable Eddington factor $f(u, -du/d\tau)$
multiplied by 3, plotted as a function of the four velocity $u_0$ ($x$ axis)
and the {\it logarithmic} four-velocity gradient $-d\ln u/d\tau|_0$ ($y$ axis).
In the relativistic regime,
the Eddington factor does depend 
only on the logarithmic velocity gradient.
}
\end{figure}

%%%%%%%%%%%%%  CONCLUDING REMARKS  %%%%%%%%%%%%%%%%%%%%%%%%%%%%
\section{Discussion}

Moment equations for relativistic radiation transfer
have been studied extensively in literature
(Lindquist 1966; Anderson \& Spiegel 1972; Hsieh \& Spiegel 1976;
Thorne 1981; Thorne et al. 1981; Udey \& Israel 1982; Schweizer 1982;
Flammang 1982, 1984; Nobili et al. 1991, 1993; Park 2001, 2006
for spherically symmetric problems;
Takahashi 2007 for the Kerr metric).
As the moment expansion gives rise to an infinite set of equations,
one must truncate it at a finite order
by adopting a suitable closure relation,
in order to make the transfer problem solvable.
When truncating at the second order, for example,
in an optically thick flow the Eddington approximation
has been usually adopted as a closure relation.

Such radiation moment formalism is quite convenient,
and it is a powerful tool for tackling problems
of relativistic radiation hydrodynamics.
However, its validity is never known
unless fully angle-dependent radiation transfer equation
is solved.

Indeed, in the relativistic regime,
where the velocity gradient becomes large
and Doppler and aberration effects become important,
even in the optically thick regime
the isotropy of the radiation field may break down.
In such a situation
the Eddington approximation can not be adopted in the comoving frame
(Fukue 2005;
see also Fukue 2008).

Instead of the usual Eddington approximation,
we can adopt a {\it variable Eddington factor},
which depends on the flow velocity $v$ and the velocity gradient $dv/d\tau$
 as well as the optical depth $\tau$
(Fukue 2006; Fukue \& Akizuki 2006, 2007; Akizuki \& Fukue 2008;
Koizumi \& Umemura 2008).
In one-dimensional flows
the variable Eddington factor $f(\tau, dv/d\tau)$ 
is generally defined as
\begin{equation}
   P_{\rm co} = f(\tau, dv/d\tau) E_{\rm co},
\label{PcoEco}
\end{equation}
where $E_{\rm co}$ and $P_{\rm co}$ are the radiation energy density
and the radiation pressure in the comoving frame, respectively.
%The closure relation in the inertial frame
%for one-dimensional flows then becomes
%\begin{equation}
%   cP(1-f\beta^2) = cE(f-\beta^2) + 2F\beta(1-f),
%\label{PEF}
%\end{equation}
%where $E$, $F$, and $P$ are the radiation energy density,
%the radiative flux, and the radiation pressure
%in the inertial frame, respectively
%(Kato et al. 1998, 2008).
In recent studies
(Fukue 2006; Fukue \& Akizuki 2006, 2007; Akizuki \& Fukue 2008;
Koizumi \& Umemura 2008),
the dependence of the variable Eddington factor
on the flow velocity has been considered.

Fukue (2008) has firstly examined the dependence 
of the variable Eddington factor on the velocity gradient
in the non-relativistic regime $\beta \leq 0.1$, obtaining:
\begin{equation}
   f \left(\tau, \frac{d\beta}{d\tau} \right)
   = \frac{1}{3} \left( 1 + \frac{16}{15} \frac{d\beta}{d\tau} \right).
\label{f_paper1}
\end{equation}

In this paper, using a similar approach,
we have numerically obtained the Eddington factor
in the relativistic regime:
\begin{equation}
   f \left(\tau, \frac{du}{d\tau} \right)
   = \frac{1}{3} \exp \left( \frac{1}{u}\frac{du}{d\tau} \right).
\label{f_paper2}
\end{equation}

Unfortunately, this relation (\ref{f_paper2}) holds only
in the relativistic regime and
does not reduce to the relation (\ref{f_paper1})
in the low speed limit of $\beta \rightarrow 0$.
However, we can easily find the bridging formula,
that reduces to equation (\ref{f_paper1}) in the non-relativistic limit
and approaches equation (\ref{f_paper2}) in the highly relativistic limit, as
\begin{equation}
   f %\left(\tau, u, \frac{du}{d\tau} \right)
   = \frac{1}{3} \exp \left( \frac{16}{15}
         \frac{\beta}{u}\frac{du}{d\tau} \right).
\label{f_paper3}
\end{equation}

Thus, the relativistic Eddington factor in 
an optically thick, plane-parallel flow
linearly decreases as a function of the velocity gradient
in the non-relativistic regime, and then
exponentially decreases as a function of 
the logarithmic velocity gradient
in the highly relativistic regime.

\section{Concluding Remarks}

In this paper, 
we have derived the Eddington factor in a plane-parallel,
relativistic and optically thick scattering medium
that is accelerating in the vertical direction;
we have introduced the one-tau photo-oval observed by the comoving observer,
and then calculated the comoving radiation field and the Eddington factor.
We thus have shown that
the Eddington factor depends on
the {\it logarithmic velocity gradient}
in the relativistic regime (\ref{f}).
Combining this result with the analytical expression of paper I,
we derived a suitable expression for the Eddington factor (\ref{f_paper3}),
valid in both the non-relativistic and fully relativistic regimes.

In order to treat the problem in the simplest form,
we only considered the optically thick regime,
where the one-tau photo-oval always exists.
In the optically thin regime or 
for large velocity gradients,
a photo-vessel may form,
where the optical depth becomes less than unity
in the downstream direction.
Furthermore, in order to isolate the effect of redshift,
we have assumed that the radiative intensity emitted from 
the one-tau photo-oval surface is constant.
In the optically thin regime or
in the extremely relativistic regime,
the comoving intensity
would be a quickly varing function of the optical depth.
This would affect the behavior of the Eddington factor
(cf. Fukue 2006; Koizumi \& Umemura 2008).
We will examine these general cases
in the future.

Under the present plane-parallel case,
the velocity increment in the downstream direction causes
the density to decrease in the same direction;
this effect is essential for the shape of 
the one-tau photo-oval and vessel.
In the spherical flow, on the other hand,
there exists a geometrical dilution effect,
and the situation can be somewhat different,
as $r^2$ terms appear (e.g., Peraiah 2002).
The present approach, however, can be 
extended to treat also the spherical case
and we will examine it in future work.

The treatment of the relativistic Eddington factor
presented in this paper may turn out to be applicable
in various aspects of relativistic astrophysics with radiation transfer;
i.e., 
black-hole accretion flows with supercritical accretion rates,
relativistic jets and winds driven by luminous central objects,
relativistic explosions including gamma-ray bursts,
neutrino transfers in supernova explosions,
and various events occured in the proto universe.

For example,
optically thick, spherically symmetric accretion flows onto black holes
were examined
(e.g., Tamazawa et al. 1975; Begelman 1978; Vitello 1978;
Burger \& Katz 1980; Flammang 1982; Park \& Miller 1991; Nobili et al. 1991),
while
spherically symmetric, relativistic radiation hydrodynamical winds
have investigated by several researchers
(e.g., Castor 1972; Ruggles \& Bath 1979; Mihalas 1980;
Quinn \& Paczy\'nski 1985; Paczy\'nski 1986; 
Turolla et al. 1986; Paczy\'nski 1990;
Nobili et al. 1994; Akizuki \& Fukue 2008).
Optically thick, supercritical accretion flow was also extensively studied
under the one-zone steady approximation
(e.g., Abramowicz et al. 1988; Szuszkiewicz et al. 1996; Beloborodov 1988).
These models have been applied to neutron-star winds
and gamma-ray bursts,
or black-hole accretion flows.
As a closure relation,
in some of these studies 
(e.g, Quinn \& Paczy\'nski 1985; Paczy\'nski 1986, 1990),
the diffusion approximation in the comoving frame and LTE was assumed.
However, the diffusion approximation may violate
in the extreme relativistic limit with large velocity gradients.
In some cases (e.g., Tamazawa et al. 1975; Turolla et al. 1986; 
Park \& Miller 1991;
Nobili et al. 1994), 
the variable Eddington factor was used.
However, the adopted Eddington factor
did only depend on the optical depth, but did not depend
on the velocity gradient.

For the problem of of the black-hole accretion flows and winds 
in the supercritical accretion regime,
radiation hydrodynamical equations were solved numerically
(e.g., Eggum et al. 1985, 1988; Kley 1989; Okuda et al. 1997;
Kley \& Lin 1999; Okuda \& Fujita 2000; Okuda 2002; Okuda et al. 2005;
Ohsuga et al. 2005; Ohsuga 2006).
These models have been applied to relativistic astrophysical jets and winds.
In these studies, however, 
simple assumptions have been made for the treatment of the radiation field;
e.g., equilibrium between gas and radiation was assumed,
the flux-limited diffusion approximation was adopted,
the flow velocity was on the order of $0.1c$,
or the optically thick to thin transition was not properly treated.
Up to now, in relation to accretion-disk winds,
no one solved the fully relativistic radiation hydrodynamical equations.

The treatment of the relativistic Eddington factor
presented in this paper will help in investigating
these types of astrophysical problems
that involve the solution of the fully relativistic
radiation hydrodynamical equations.

\vspace*{1pc}

The author would like to express his sincere thanks to M. Umemura,
T. Koizumi, and C. Akizuki
for their enlightening and stimulated discussions.
Valuable comments to improve the original manuscript
 from anonymous referees are also gratefully acknowledged.
This work has been supported in part
by a Grant-in-Aid for Scientific Research (18540240 J.F.) 
of the Ministry of Education, Culture, Sports, Science and Technology.

%\noindent

\end{document}